\documentclass[fleqn,usenatbib]{mnras}
 
\usepackage{graphicx} 
\usepackage{amsmath}
\usepackage{booktabs}
\usepackage{mathtools}
\usepackage{xcolor}

\usepackage{txfonts} 
\usepackage{natbib}               
\title[]{Spectro-Photometry and Radial Distribution of Multiple Stellar Populations in Globular Clusters from Gaia XP Spectra}
\author[V.\,J.\,Mehta et al.] 
       {
       V.\,J.\,Mehta$^{1}$,
       A.\,P.\,Milone$^{2,3}$, 
       L.\,Casagrande$^{1}$,
       A.\,F.\,Marino$^{3}$, 
       M.\,V.\,Legnardi$^{2}$, 
       G.\,Cordoni$^{1}$, 
       E.\,Dondoglio$^{3}$, 
          \newauthor 
          S.\,Jang$^{4}$,
          S.\,Lionetto$^{2}$,      
          T.\,Ziliotto$^{2}$,
          M.\,Barbieri$^{2}$,
          M.\,Bernizzoni$^{2}$,
          E.\,Bortolan$^{2}$,
          A.\,Bouras Moreno Sanchez$^{2}$,
          \newauthor
      E.\,P.\,Lagioia$^{5}$,
              A.\,Mohandasan$^{2}$,
          F.\,Muratore$^{2}$
        \\
        $^{1}$ Research School of Astronomy and Astrophysics, Australian National University, Canberra, ACT 2611, Australia\\
$^{2}$Dipartimento di Fisica e Astronomia ``Galileo Galilei'', Univ. di Padova, Vicolo dell'Osservatorio 3, Padova, IT-35122\\
$^{3}$Istituto Nazionale di Astrofisica - Osservatorio Astronomico di Padova, Vicolo dell'Osservatorio 5, Padova, IT-35122\\
$^{4}$Center for Galaxy Evolution Research and Department of Astronomy, Yonsei University, Seoul 03722, Korea \\
$^{5}$South-Western Institute for Astronomy Research, Yunnan University, Kunming, 650500 P. R. China
}
\begin{document}

\maketitle 
\label{firstpage}
\begin{abstract}
  Understanding the formation of multiple populations in globular clusters (GCs) represents a challenge for stellar population studies. Nevertheless, the outermost cluster regions, likely to hold clues about the initial configuration of GC stars, remain underexplored.
We use synthetic spectra reflecting the chemical compositions of first- and second-population (1P, 2P) stars in 47\,Tucanae to identify spectral regions sensitive to these populations. This led us to define new photometric bands that effectively distinguish 1P and 2P giant stars using Gaia XP spectra.
Testing these filters, we constructed  the pseudo two-color diagrams dubbed chromosome maps (ChMs) and, for the first time, identified 1P and 2P stars in the cluster's outermost regions and beyond its tidal radius.
 We constructed similar diagrams for NGC\,3201, NGC\,6121, NGC\,6752, and NGC\,6397, thus exploring GCs with different metallicities. The ChMs effectively distinguished multiple populations in the outer regions of all clusters, except for the metal-poor NGC 6397.  Our findings, together with literature results from more-internal regions, show that the 2P stars of 47\,Tucanae are more-centrally concentrated than the 1P.  A similar pattern is seen for 2P stars with extreme chemical composition of NGC\,3201. The multiple populations of NGC\,6121, and NGC\,6752 share the same radial distributions. These radial behaviors are consistent with the GC formation scenarios where 2P stars originate in the central regions. Noticeably, results on NGC\,3201 are in tension with the conclusion from recent work that its 1P is more centrally concentrated than the 2P and might form with more central concentration.
\end{abstract}


\section{Introduction}\label{sec:intro}

Globular clusters (GCs) with initial masses larger than $\sim 10^{5} M_{\odot}$ contain two main stellar groups. The first population (1P) exhibits a chemical composition comparable to that of  most Milky Way stars with comparable metal content. 
  Conversely, the second population (2P) includes stars enhanced in He, N, and Na, and depleted in C and O \citep[e.g.][]{kraft1994a, gratton2012a, bastian2018a, milone2022a}. 
  Moreover, some GCs (hereafter type II GCs) host 
    stars which also differ in their heavy-element content \citep[e.g.][]{yong2008a, marino2009a, marino2015a, milone2017a, johnson2015a}.

Despite extensive observational and theoretical efforts, the mechanisms responsible for the formation of multiple stellar populations and the contributions of GCs to the early Galaxy's assembly remain poorly understood.
Various authors propose that the presence of multiple populations is linked to distinct star-formation events. In their scenarios, the 2P stars formed from material contaminated by more massive 1P stars, and the stars lost by the GCs provided a significant contribution to the Galactic halo \citep{dantona1983a, dantona2016a, decressin2007a, denissenkov2014a, renzini2022a, lacchin2024a}.
 Alternatively, other authors suggest that 2P stars may arise either from  binary-star evolution, or accumulating enriched material onto pre-existing 1P stars. In these scenarios, GCs provided moderate contribution to the Galactic halo assembly \citep[e.g.][]{wang2020a, gieles2018a}.

The current radial behavior of 1P and 2P stars offers valuable insights into the origins and evolution of multiple populations. Specifically, GCs that have not fully undergone dynamic relaxation may preserve information regarding the original distribution of these stellar populations  \citep[e.g.][]{vesperini2013a, vesperini2018a, vesperini2021a, brunet2015a, dalessandro2019a, sollima2021a}. Hence, it is crucial to study the multiple populations in  the outermost cluster regions, which are characterized by long relaxation times.

Photometry proves to be an effective technique for exploring multiple stellar populations. 
 The 1P and 2P stars form distinct sequences in appropriate photometric diagrams that are sensitive to their light-element abundances, thus providing the opportunity to 
  analyze these populations in large stellar samples.

 Nowadays, photometric surveys of multiple populations have been performed by using Hubble Space Telescope (HST) observations of 
  both main-sequence (MS) stars and of evolved stars belonging to the red-giant, horizontal, and asymptotic-giant branches 
  \citep[RGB, HB, AGB, e.g.\,][]{milone2017a, milone2020a,  lagioia2021a, dondoglio2021a}. 
    These works, limited by the narrow field of view of  HST, have primarily explored the central regions of the clusters.

 Wide-field ground based photometry has been complementary to HST studies since it is instrumental to detect multiple stellar populations in the external GC regions \citep{ marino2008a, yong2008b, milone2012b, monelli2013a, jang2022a, lee2019a, lee2022a}. Nevertheless, these studies rarely approach the GC tidal radius while the cluster outskirts are nearly unexplored  \citep[but see][for recent photometric investigations of multiple populations beyond the tidal radius of NGC\,1851]{dondoglio2023a}.
 
The Gaia Data Release 3 provides the unique opportunity to explore multiple populations in the outermost GC regions.
Recently, \citet{cordoni2023a} exploited the low-resolution XP spectra from Gaia DR3 to examine NGC 1851, a Type II GC. By using these spectra, they derived photometry in various bands, including the B, V, I, and the f415$^{25}$  filter, introduced in their work. The resulting photometric diagrams allowed them to detect the stellar populations with distinct heavy-element chemical compositions in the outskirts of the cluster, even beyond its tidal radius.

In this work, we focus on Type\,I GCs, which are characterized by stellar populations with similar metallicities but different contents of some light elements, including He, C, N, and O.
 The structure of the manuscript is as follows:
In Section\,\ref{sec:data}, we present our dataset.
Section\,\ref{sec:47t} exploits simulated spectra of 1P and 2P stars to define the most effective photometric bands to 
  untangle stars with different elemental abundances in 47\,Tucanae. These filters are used  to construct photometric diagrams that allow us to  detect 1P and 2P stars in GCs  based on low-resolution XP spectra from Gaia Data Release 3  \citep[DR3,][]{gaia2023a}. We first adopted 47\,Tucanae as a test case, and then extended the analysis to the GCs NGC\,6121, NGC\,3201, NGC\,6752, and NGC\,6397.
 The radial distribution of the multiple stellar populations is discussed in Section\,\ref{sec:RD}, while  Section\,\ref{sec:summary} offers a summary and discussion.

\section{Data}\label{sec:data}
The primary dataset we relied on to explore the GCs 
 consists in low-resolution photometry, spectroscopy, astrometry, and proper motions,  all sourced from Gaia DR3.

Gaia DR3 offers spectra with resolutions of about 30 to 100 for $\sim$200 million astronomical sources. Each spectrum, known as XP spectrum, comprises two distinct spectra obtained by the two components of the Gaia low resolution spectro-photometer. These spectra, 
  termed 
  'blue photometer' (BP) and 'red photometer' (RP)  cover the spectral intervals of 3300–6800 \AA\,and 6400–10500 \AA, respectively. 
 
 We used these spectra, which are accessible through the Gaia archive in the form of two sets of 55 Hermite function coefficients for each channel (BP$-$RP) to derive stellar photometry. The spectra are transformed into standard calibrated spectra, represented as flux versus wavelength, using the publicly available Python library GaiaXPy\footnote{https://gaia-dpci.github.io/GaiaXPy-website/}.

We used the stellar proper motions and parallaxes along with the parameters provided by Gaia DR3 that are indicative of the photometric and astrometric quality   to identify a group of accurately measured cluster members.
 To do this, we followed the recipe from our previous works \citep{milone2018a, cordoni2023a}.
Moreover, we accounted for the contamination from nearby sources by using the
BP/BR Excess Factor, $\epsilon$, which is a measure of the ratio between the sum of BP and RP, and G magnitudes and indicates the consistency of Gaia photometry. To select poorly-contaminated stars, the following constraint is applied \citep{gaia2021a}:

\begin{align}
\begin{split}
	0.01+0.039(\mathrm{BP}-\mathrm{RP})&<\log_{10}\epsilon \\
	&<0.12+0.039(\mathrm{BP}-\mathrm{RP})
	\label{eq:exc}
\end{split}
\end{align}

\begin{figure*} 
    \includegraphics[width=14.5cm,trim={0.5cm 7cm 1.5cm 5.0cm},clip]{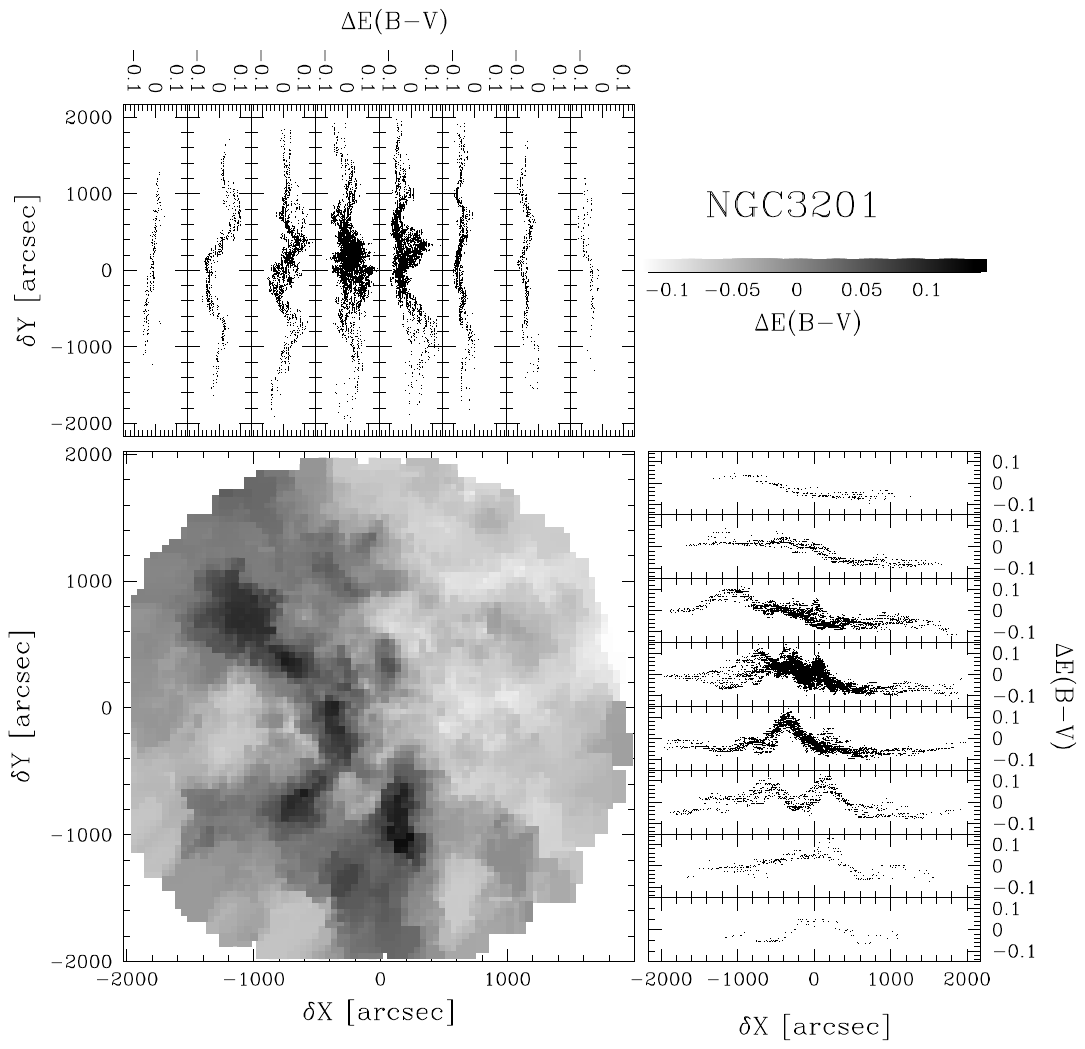}
    \caption{Differential-reddening map in the direction of NGC\,3201 (bottom left).  The x-axis and y-axis are aligned with the directions of right ascension and declination, respectively, with the coordinates, $\delta$X and $\delta$Y, given relative to the cluster center. The top-left and bottom-right panels illustrate the reddening variation, $\Delta$E(B$-$V), relative to the average reddening of the analyzed stars, plotted as a function of the $\delta$X and $\delta$Y coordinates for stars in eight slices of $\delta$Y and $\delta$X, respectively.}
    \label{fig:maps}
\end{figure*}

Photometry of cluster members in NGC\,6121, NGC\,3201, and NGC\,6397 has been corrected for differential reddening through the approach outlined by \citet[][see their Section 2.1]{milone2012a}. 
To summarize, we calculated the ridge line for  MS, sub-giant branch (SGB), and RGB stars in the $G_{\rm RP}$ vs. $G_{\rm BP}-G_{\rm RP}$ color-magnitude diagram (CMD) and  used it a reference to derive the color residuals.
To assess the extent of differential reddening experienced by each star in the examined field of view, we assembled a sample of 35 neighbors consisting of bright MS, SGB, and faint-RGB cluster members without apparent binary features.
 Our most accurate estimate of differential reddening is determined by the median of the color residuals, which are computed along the direction of the reddening vector.
The reddening direction is derived by using appropriate absorption coefficients that we calculated as in \citet{legnardi2023a}.
We adopted the reddening laws corresponding to three values of the ratio, $R_{\rm V}$, between the absorption in the $V$ band, $A_{\rm V}$, and the difference between the absorption in the B and V bands, E(B$-$V). For NGC\,6397, we assumed the canonical value, R$_{\rm V}$=3.1, whereas for NGC\,6121 and NGC\,3201 we adopted R$_{\rm V}$=3.85 and R$_{\rm V}$=3.19, respectively, as inferred by \cite{legnardi2023a} from multi-band HST photometry.
 As an example, we show the differential-reddening map in the direction of NGC\,3201 in Figure\,\ref{fig:maps}.

\section{Hunting multiple populations in GCs with Gaia XP spectra}\label{sec:47t}

 To detect the distinct stellar populations, we defined a set of filters that maximize the separation of the 1P and 2P star sequences in the GC CMDs. To do this, we first adopted 47\,Tucanae as a test case (Section\,\ref{subsec:47t}), and then investigate  more metal-poor GCs, namely NGC\,6121, NGC\,6752, NGC\,3201, and NGC\,6397 (Section\,\ref{subsec:MPs}). This allowed an exploration of the efficiency of our method from [Fe/H]$\sim -0.75$ to [Fe/H]$\sim -2.02$ dex.

\subsection{Multiple populations in 47 Tucanae}\label{subsec:47t}
We used two synthetic spectra with [Fe/H]=$-$0.75 and [$\alpha$/Fe]=0.3 that represent two stars on the bottom of the RGB of 47\,Tucanae. Both spectra have an effective temperature T$_{\rm eff}$=4746 K, and a gravity, $\log{g}=1.82$ but differ in their contents of some light elements. The first spectrum has [C/Fe]=0.0, [N/Fe]=0.0 and [O/Fe]=0.3, similar with what is observed in 1P stars. The other spectrum represents a 2P star with extreme chemical composition and is enhanced in [N/Fe] by 1.2 dex and depleted in both [C/Fe] and [O/Fe] by 0.5 dex with respect to the 1P spectrum. The spectra are derived by \cite{milone2023a} by using the the SYNTHE programme \citep{kurucz1981a, castelli2005a, kurucz2005a} and the model atmospheres that they calculated with the ATLAS 12 computer program \citep{kurucz1970a, kurucz1993a, sbordone2004a}.

\begin{figure*} 
    \includegraphics[width=14.5cm,trim={0.5cm 6cm 0.5cm 12.0cm},clip]{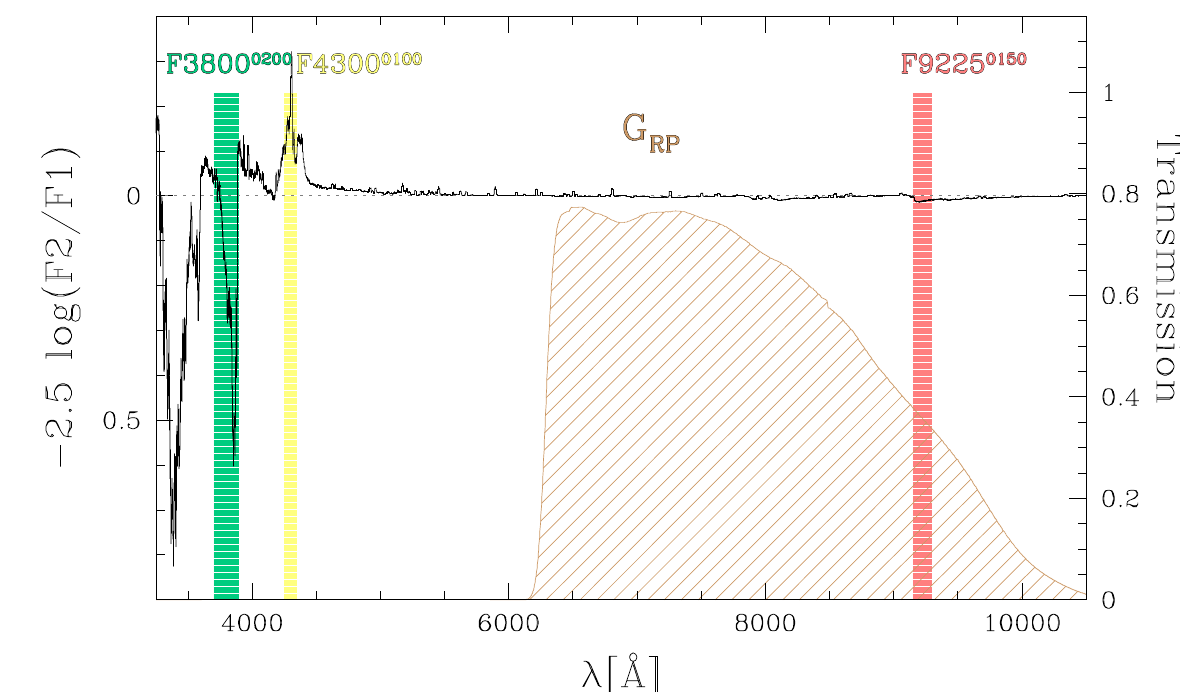}
    \caption{Flux ratio of two spectra with [Fe/H]=$-$0.75 and C, N, and O abundances indicative of 2P and 1P stars \citep{milone2023a}. The transmission curve of the Gaia DR3 G$\rm RP$ filter is colored brown, whereas the transmission curves of the $m3800^{0200}$, $m4300^{0100}$, and $m9225^{0150}$ bands are represented by the aqua, yellow, and red shaded areas, respectively.}
    \label{fig:filtri}
\end{figure*}

 To illustrate the differences between the simulated spectra, we 
  show the 2P/1P flux ratio, in logarithmic scale, 
  against  the wavelength, $\lambda$, in Figure\,\ref{fig:filtri}. As expected, the most significant variations involve the regions around 3,400\AA\, and 3,800\AA. Here, the 2P spectrum is more absorbed than the 1P one, owing to the stronger NH and CN molecular bands. In addition, around 4,300 \AA\, the 2P spectrum is less absorbed than the 1P one, due to the weaker CH bands.

 To find the bands that are effective in photometrically distinguishing between 1P and 2P stars we defined a grid of magnitudes as:

\begin{equation}
    m_{\lambda_{0}}^{\delta \lambda} =-2.5\log\frac{\int \lambda f_\lambda T (\lambda_{0},\delta \lambda)\mathrm{d}\lambda}{\int \lambda T ({\lambda_{0},\delta \lambda)} \mathrm{d}\lambda}  
    \label{eq:mag}
\end{equation}

where $\lambda$ and $f_\lambda$ are the wavelengths and fluxes (in units $\mathrm{erg s^{-1} cm^{-2} A^{-1}}$), respectively, of the star and $T$ refers to the transmission curve of the bandpass being used to derive magnitudes. We adopted transmission curves defined by step functions with center $\lambda_{0}$ and width $\delta \lambda$:

 \begin{align}
     T (\lambda_{0},\delta \lambda) = 
     \begin{dcases*}
        1, & \text{if } $\lambda_{0}-\delta \lambda /2 <\lambda \leq \lambda_{0}+\delta \lambda /2 $,\\
        0, & \text{otherwise.}
        \end{dcases*}
  \end{align}

We calculated grids of magnitudes by varying the value of $\lambda_{0}$ from 3,200 to 10,000 \AA\, and by assuming four different values of $\delta \lambda=100, 150, 200,$ and 1,000 \AA.
 For each combination of $\lambda_{0}$ and $\delta \lambda$, we calculated the magnitude difference between 1P and 2P stars. We found that the most-efficient filters to distinguish the distinct stellar populations of 47\,Tucanae are F3800$^{0200}$ and F4300$^{0100}$, which are sensitive to N and C variations, respectively.
 In particular, the pseudo color $C_{\rm m3800, m4300, GRP}$=($m3800^{0200}-m4300^{0100}$)$-$($m4300^{0100}-G_{\rm RP}$)  maximizes the separation between 1P and 2P stars. Indeed,  for a fixed luminosity, the 2P stars, which are 
 N-rich and C-poor,  exhibit smaller ($m4300^{0100}-G_{\rm RP}$) and larger ($m3800^{0200}-m4300^{0100}$) colors compared to the 1P stars.
  Magnitudes constructed with filters centered around $\lambda_{0} \sim 3,400$\,AA, such as F3400$^{0100}$, would be very sensitive to stellar populations with different nitrogen abundances. Nevertheless, we excluded photometry in these bands from the analysis. Indeed, as already demonstrated by \citet{cordoni2023a}, we verified that it is not possible to derive accurate determinations of the UV stellar magnitudes from the Gaia XP spectra, due to the poorer quality of the spectra in the extreme UV region.

 Moreover, we notice that the 2P stars are hotter than 1P stars with the same luminosity because of their higher helium content. Hence, they exhibit bluer colors than 1P stars.
 As a consequence, a wide color baseline, like $m4300^{0100}-m9225^{0150}$ would further increase the separation in the CMD between the sequences of 1P and 2P stars with different C and He abundances.

\begin{figure*}
    \includegraphics[width=8.5cm,trim={0.5cm 5cm 0.5cm 4.5cm},clip]{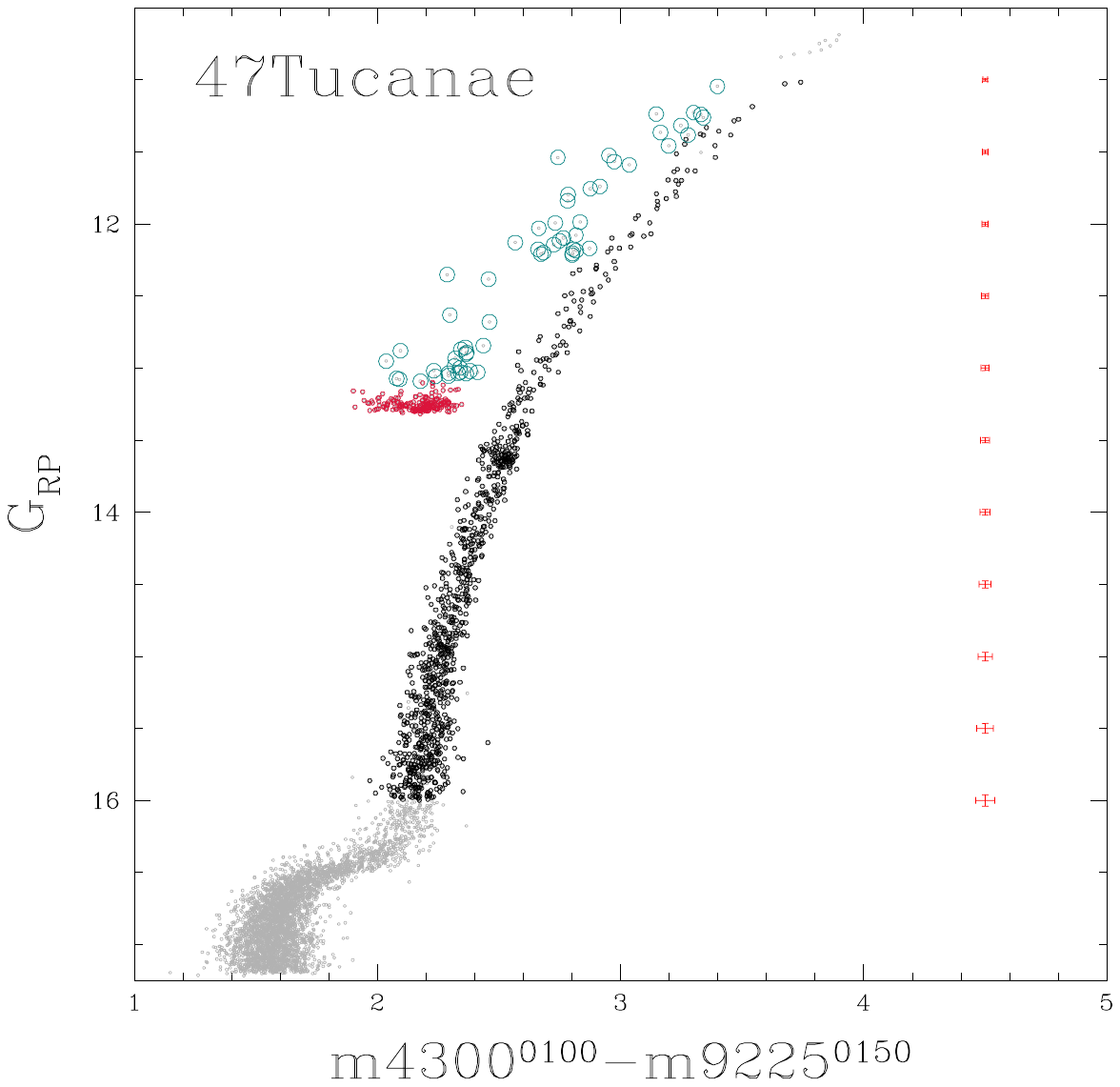}
    \includegraphics[width=8.5cm,trim={0.5cm 5cm 0.5cm 4.5cm},clip]{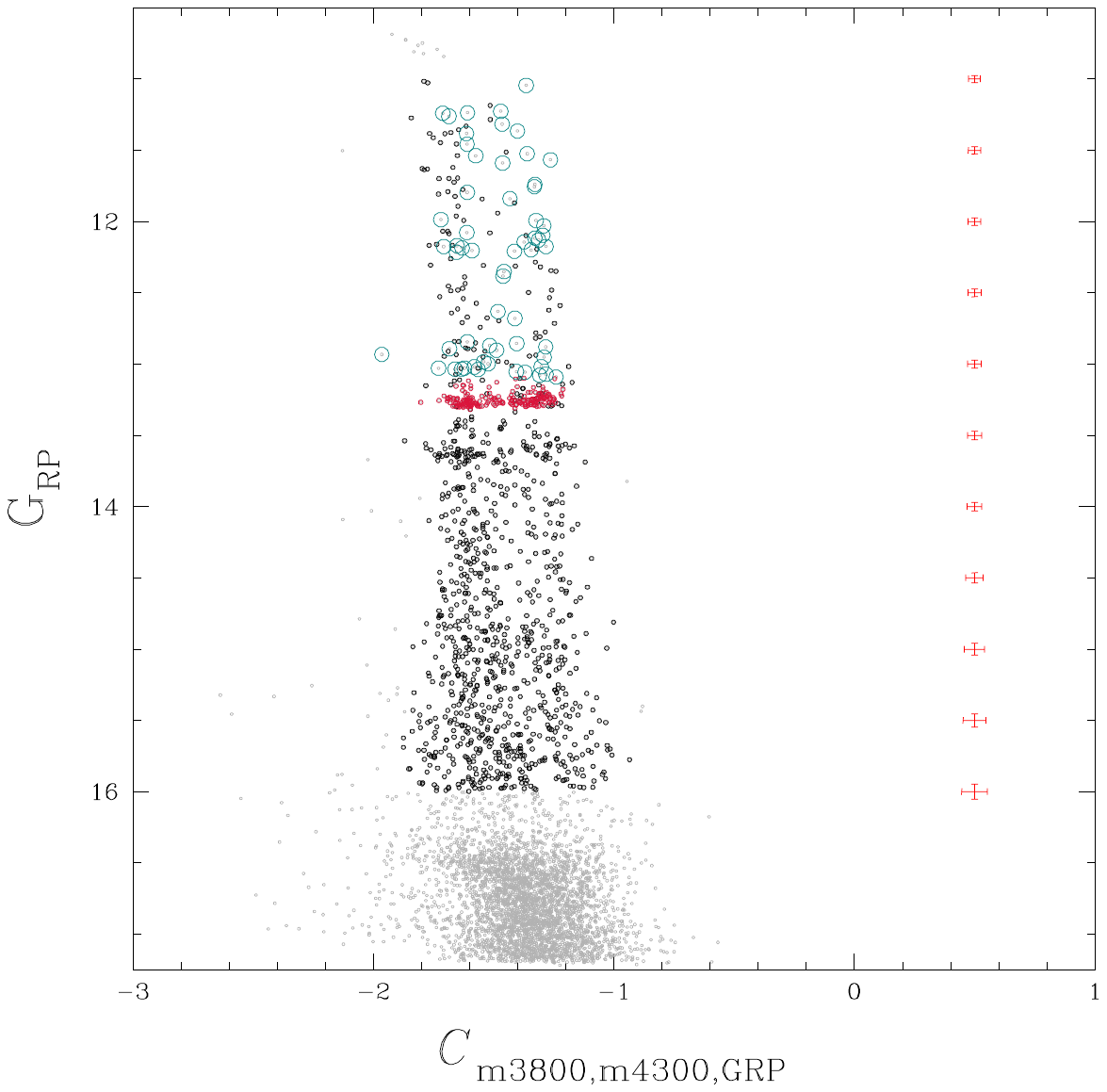}
    \caption{G$_{\rm RP}$ vs.\,$m4300^{0100}-m9225^{0150}$ CMD (left) and G$_{\rm RP}$ vs.\,$C_{\rm m3800, m4300, GRP}$ pseudo-CMD (right) for proper-motion selected stars of 47\,Tucanae. Bona-fide RGB, red HB, and AGB stars are colored black, crimson, and teal, respectively. }
    \label{fig:47tCMD}
\end{figure*}

The resulting G$_{\rm RP}$ vs.\,$m4300^{0100}-m9225^{0150}$ and G$_{\rm RP}$ vs.\,$C_{\rm m3800, m4300, GRP}$ diagrams for the probable cluster members of 47\,Tucanae are plotted in Figure\,\ref{fig:47tCMD}, where we used red and teal colors to represent the probable red HB and AGB stars that we selected by eye. 
The evolutionary sequences of both diagrams show broader color spreads than those expected from observational uncertainties alone, a characteristic expected for GCs with multiple populations. Notably, distinct features such as split RGB, red HB, and AGB sequences are discernible in the G$_{\rm RP}$ vs.\,$C_{\rm m3800, m4300, GRP}$ pseudo-CMD. For clarity, the RGB stars where the multiple sequences are more distinctly visible are highlighted in black.

\begin{figure*}
    \includegraphics[width=8.5cm,trim={0.5cm 5cm 0.5cm 4.5cm},clip]{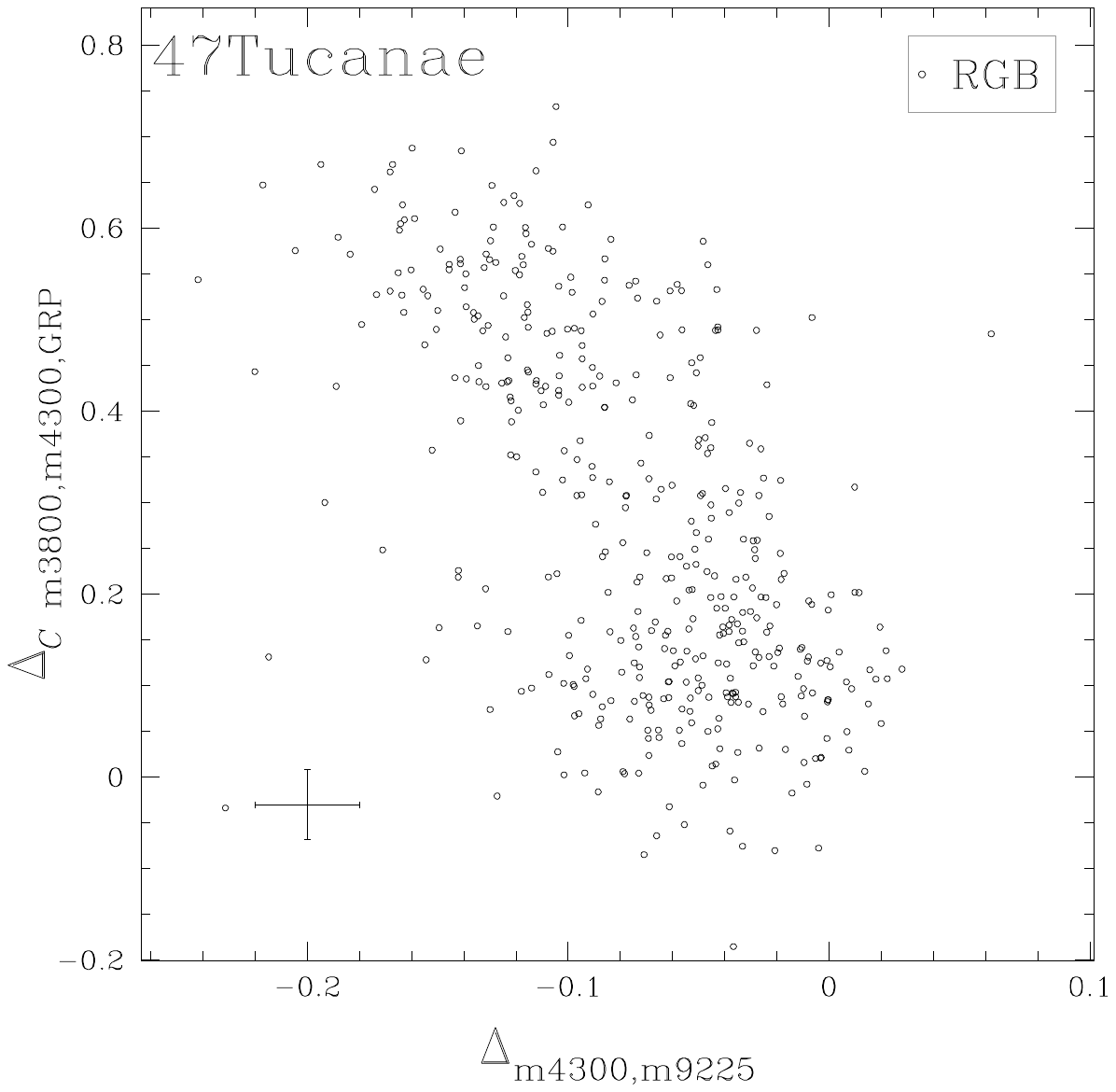}
    \includegraphics[width=8.5cm,trim={0.5cm 5cm 0.5cm 4.5cm},clip]{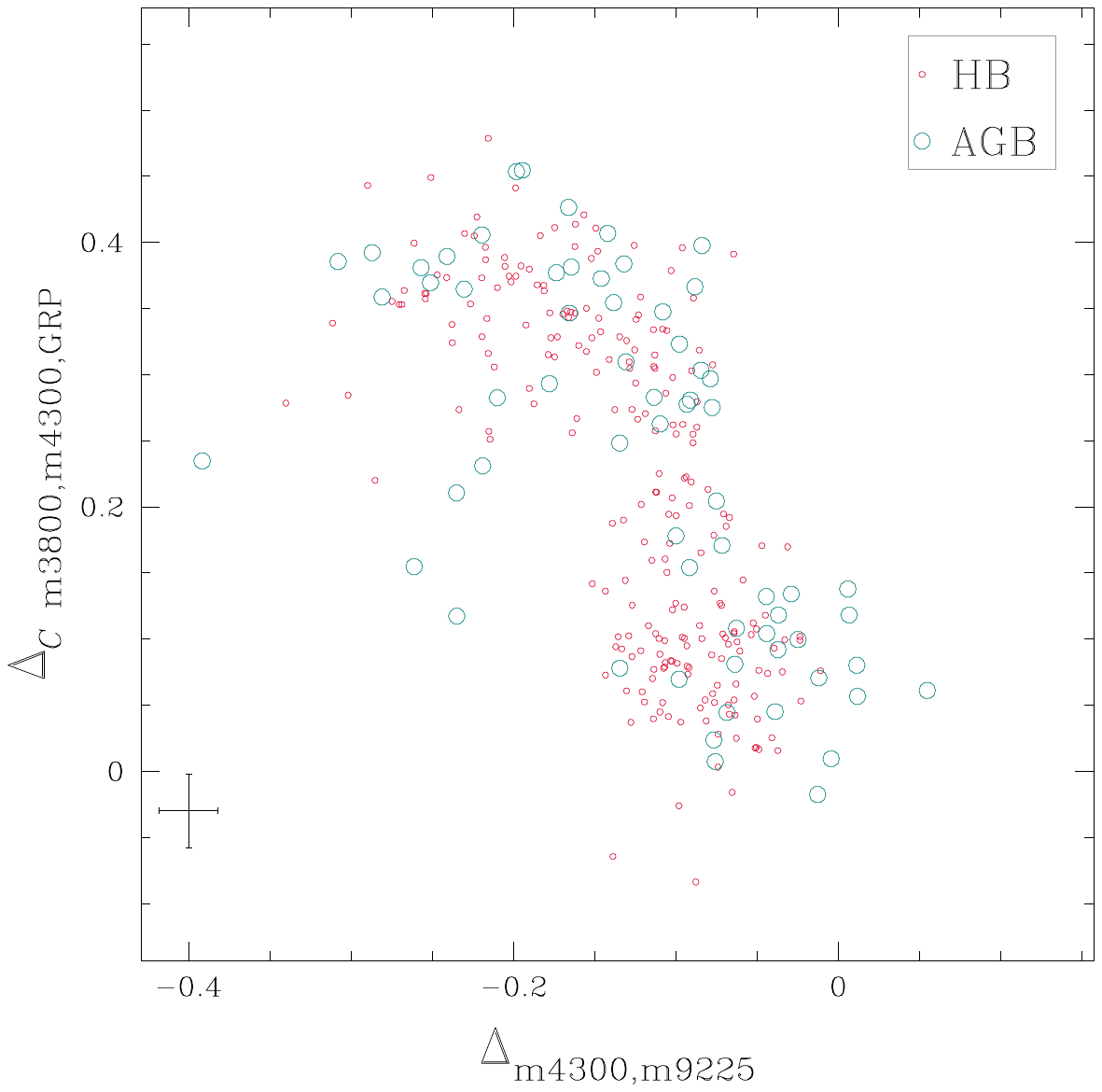}
    \caption{$\Delta_{ {\it C} \rm m3800,m4300,GRP}$ vs.\,$\Delta_{\rm m4300,m9225}$ ChM for 47\,Tucanae stars.  The left panel displays the ChM of RGB stars, whereas the HB and AGB stars are represented in the right panel with crimson and teal colors, respectively. }
    \label{fig:47tChM}
\end{figure*}

To better disentangle the multiple populations in the RGB of 47\,Tucanae, we exploited the photometric diagrams of Figure\,\ref{fig:47tCMD} to construct the pseudo two-color diagram  also known as chromosome map (ChM).
 In a nutshell, we derived 
  the red and blue boundaries of the RGB and computed the $\Delta_{\rm {\it C} m3800,m4300,GRP}$ and $\Delta_{\rm m4300,m9225}$ pseudo colors  by following the recipe by \citet[][see their equations 1 and 2]{milone2017a}. The resulting $\Delta_{\rm {\it C} m3800,m4300,GRP}$ vs.\,$\Delta_{\rm m4300,m9225}$ ChM is plotted in the left panel of Figure\,\ref{fig:47tChM}.
  The color distribution in this ChM is not uniform. About half RGB stars are  grouped near the origin of the ChM reference frame and have $\Delta_{ {\it C} \rm m3800,m4300,GRP}$ values smaller than $\sim 0.2$ mag. The remaining RGB stars are distributed along a sequence that extends up towards large absolute values of $\Delta_{\rm {\it C} m3800,m4300,GRP}$ and $\Delta_{\rm m4300,m9225}$.
  
 Similarly, we derived the ChMs of AGB and red-HB stars, which are plotted in the right panel of Figure\,\ref{fig:47tChM}. Notably, the distribution of stars across these ChMs is qualitatively similar to that of the RGB ChM.

\begin{figure*}
    \includegraphics[width=13cm,trim={0.5cm 5.5cm 0.5cm 4.5cm},clip]{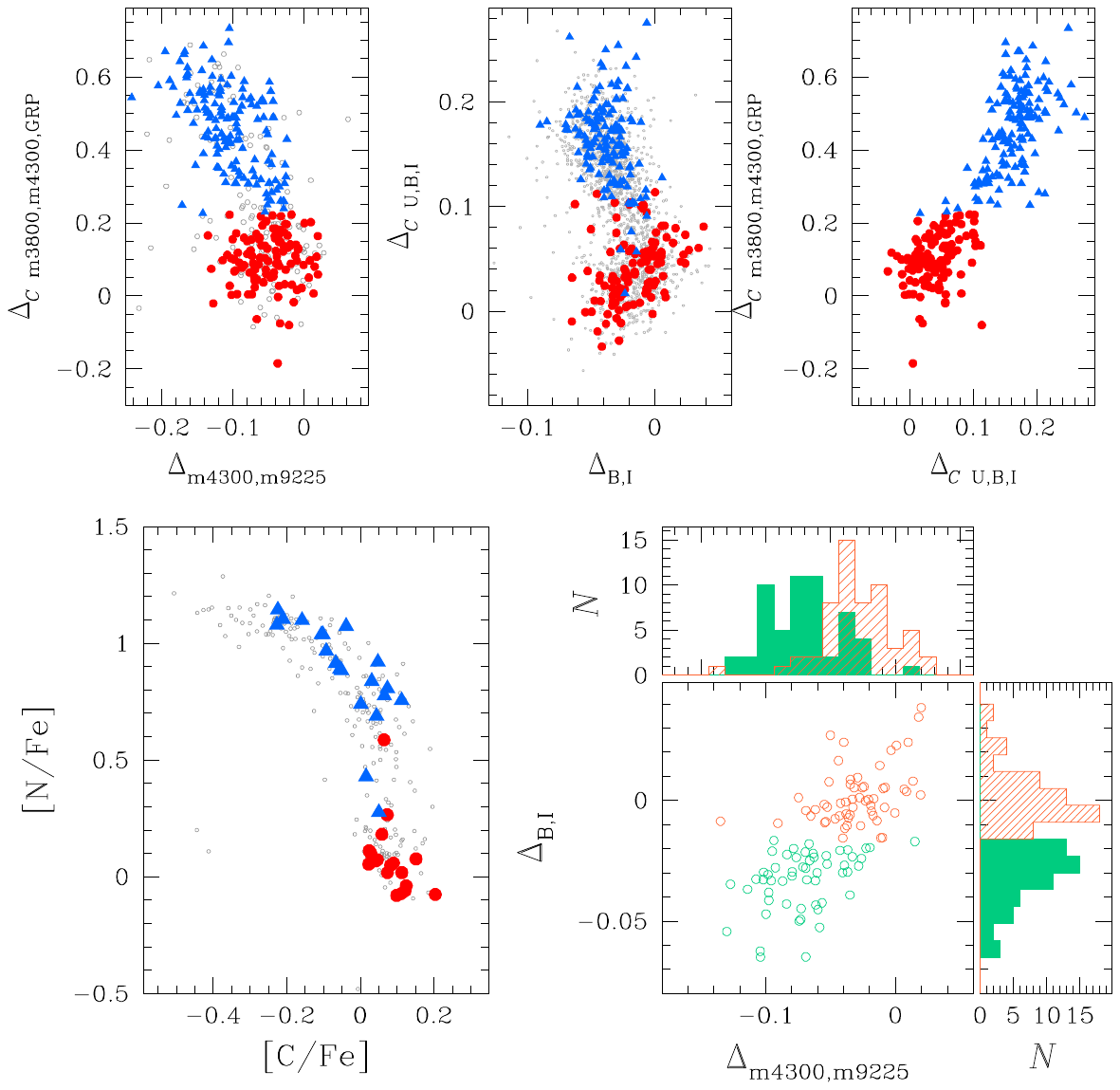}
    \caption{\textit{Top panels.} Reproduction of the $\Delta_{\rm {\it C} m3800,m4300,GRP}$ vs.\,$\Delta_{\rm m4300,m9225}$ ChM of Figure\,\ref{fig:47tChM} (left), $\Delta_{\rm {\it C} U,B,I}$ vs.\,$\Delta_{\rm B,I}$ ChM from \citet{jang2022a} (middle) and $\Delta_{\rm {\it C} m3800,m4300,GRP}$ vs.\,$\Delta_{{\it C} U,B,I}$ ChM (right). The bottom-left panel shows the nitrogen-carbon anticorrelation from APOGEE DR17. The probable 1P and 2P stars, selected from the top-left ChM and analyzed both in our work and in the papers by \citet{jang2022a} or \citet{abdurro2022a},  are represented with red dots and blue triangles, respectively. Bottom-right panels show $\Delta_{\rm B,I}$ as a function of $\Delta_{\rm m4300,m9225}$ for 1P metal-rich (orange) and metal-poor (aqua) stars. We also show the corresponding $\Delta_{\rm B,I}$ and $\Delta_{\rm m4300,m9225}$ histogram distributions.  } 
    \label{fig:47tucVAL}
\end{figure*}

Figure\,\ref{fig:47tucVAL} compares the distribution of RGB stars in the ChM of Figure\,\ref{fig:47tChM}, which is reproduced in the top-left panel, and in other photometric and spectroscopic diagrams used in the literature to disentangle 1P and 2P stars in 47\,Tucanae.
 The top-middle panel reproduces the $\Delta_{\rm {\it C} U,B,I}$ vs.\,$\Delta_{\rm B,I}$ ChM derived by \citet{jang2022a}, where the 1P stars are clustered around the origin of this ChM, whereas 2P stars have $\Delta_{\rm {\it C} U,B,I} > 0.1$ mag. The top-right panel shows $\Delta_{\rm {\it C} m3800,m4300,GRP}$ as a function of $\Delta_{\rm {\it C} U,B,I}$, thus combining photometry from our work and from \cite{jang2022a}.  In the bottom-left panel of Figure\,\ref{fig:47tucVAL} we utilized the abundances of carbon and nitrogen relative to iron from APOGEE DR17 \citep{abdurro2022a} to illustrate the anticorrelation between [C/Fe] and [N/Fe]. In this diagram, the 1P stars exhibit nearly solar abundances of carbon and nitrogen, whereas 2P stars are depleted in carbon and enhanced in nitrogen.

 We used red dots and blue triangles to represent the stars with $\Delta_{ \rm {\it C} \rm m3800,m4300,GRP}$ values smaller and larger than $\sim 0.2$ mag, respectively, for which either accurate $U,B,I$ photometry from \citep{jang2022a} or spectroscopy from APOGEE DR17 is available.  The fact that the bulk of the red stars, that we selected from the top-left ChM, populates the 1P sequence in the  $\Delta_{\rm {\it C} U,B,I}$ vs.\,$\Delta_{\rm B,I}$ ChM demonstrates that they correspond to the first population of 47\,Tucanae, whereas the selected blue stars mostly belong to the second population. This conclusion is corroborated by the correlation between the $\Delta_{\rm {\it C} m3800,m4300,GRP}$ and $\Delta_{\rm {\it C} U,B,I}$ (top-right panels) and by the fact that the bulk of the selected red stars share the same carbon and nitrogen abundances as 1P stars. In contrast, stars represented by blue triangles show depletion in carbon and enhancement in nitrogen, as expected for the 2P. 
 We conclude that the $\Delta_{\rm {\it C} m3800,m4300,GRP}$ vs.\,$\Delta_{\rm m4300,m9225}$ ChM is an effective instrument to disentangle multiple populations in 47\,Tucanae.

 Recent works have revealed an extended 1P sequence in the ChMs of 47\,Tucanae derived from photometry collected either with the Hubble and James Webb telescopes or with ground-based facilities \citep{milone2017a, jang2022a, milone2023a,  marino2024a}. The pseudo-color extension of 1P stars is due to metallicity variations among 1P stars of [Fe/H]$\sim 0.1$ dex \citep{legnardi2022a, marino2023a}. 
 The lower-right panel of Figure\,\ref{fig:47tucVAL} shows $\Delta_{\rm B,I}$
 vs.\,$\Delta_{\rm m4300,m9225}$ for the probable 1P stars selected in the top-left panel alone. We defined two equally-populated groups of stars with different $\Delta_{\rm B,I}$ values, colored them aqua and orange, and represented the $\Delta_{\rm B,I}$ and $\Delta_{\rm m4300,m9225}$ histogram distribution of each group. The correlation between $\Delta_{\rm m4300,m9225}$ and  $\Delta_{\rm B,I}$ and the finding that aqua and orange stars exhibit different average $\Delta_{\rm m4300,m9225}$ values corroborates the evidence that the color extension of the sequence is intrinsic, thus indicating the $\Delta_{\rm m4300,m9225}$ pseudo-color is an effective tool for detecting subtle metallicity variations among 1P stars.

\begin{figure}
    \includegraphics[width=9.2cm,trim={0.5cm 5.7cm 0.0cm 12.5cm},clip]{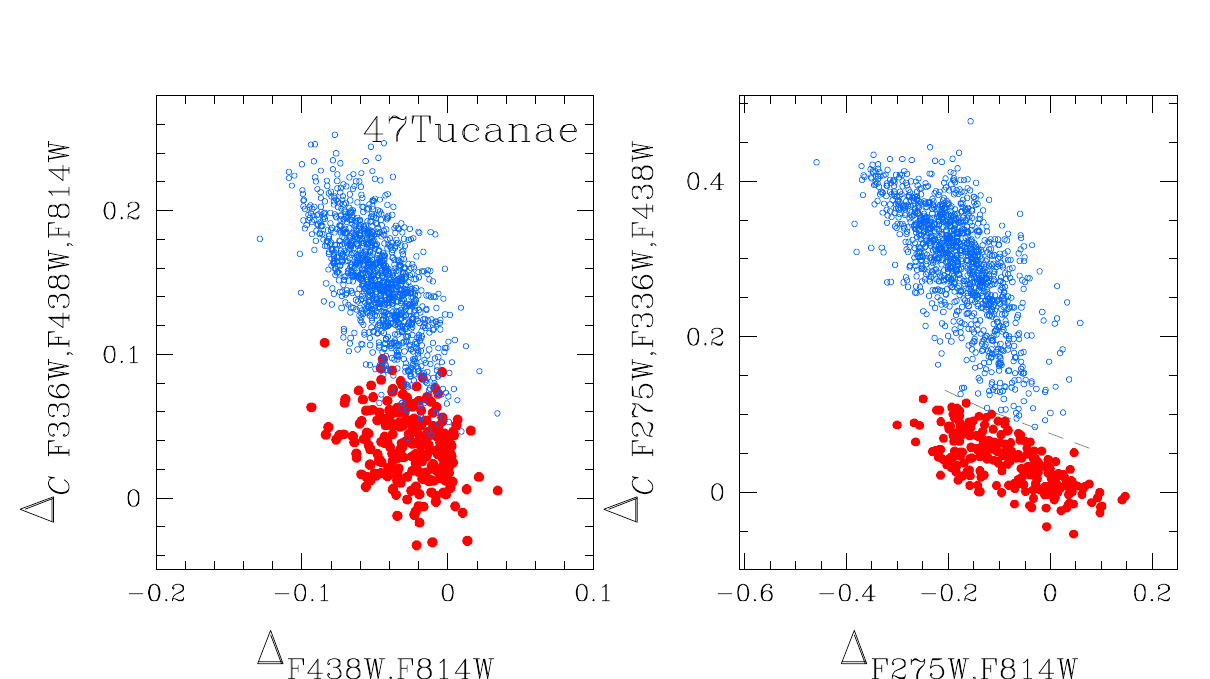}
   \caption{ Comparison between the 47\,Tucanae ChM constructed with the F336W, F438W, and F814W filters (left) and the one derived from F275W, F336W, F438W, and F814W magnitudes by \citet{milone2017a}. The dashed line separates the bulk of 1P and 2P stars, which are colored red and blue, respectively. }
    \label{fig:47tHST}
\end{figure}

Previous studies of 1P and 2P stars in the central regions of 47\,Tucanae and various Galactic GCs have utilized the $\Delta_{C \rm F275W,F336W,F438W}$ vs.,$\Delta_{\rm F275W,F814W}$ ChM \citep[e.g.,][]{milone2017a, lagioia2024a}. For completeness, the left panel of Figure \ref{fig:47tHST} presents the $\Delta_{C \rm F336W,F438W,F814W}$ vs.\,$\Delta_{\rm F438W,F814W}$ ChM of 47 Tucanae, constructed with filters similar to the Johnson-Cousins U, B, and I bands. The 1P and 2P stars selected by \citet{milone2017a} from the $\Delta_{C \rm F275W,F336W,F438W}$ vs.\,$\Delta_{\rm F275W,F814W}$ ChM (right panel of Figure \ref{fig:47tHST}) clearly occupy distinct regions in both ChMs. This demonstrates that results from U, B, I photometry for 47\,Tucanae are directly comparable with literature findings based on F275W, F336W, F438W, and F814W data.

\subsection{Multiple populations in low-metallicity GCs}\label{subsec:MPs}
In this subsection, we apply the procedure that we have introduced for 47\,Tucanae to derive photometry in the m3800$^{0100}$, m4300$^{0100}$ and m9225$^{0150}$ bands for stars in 
 three
 GCs with different [Fe/H] values: 
  NGC\,6121, NGC\,6752, 
 and NGC\,6397. According to the 2010 edition of the \citet{harris1996a} catalog, these clusters have iron abundances [Fe/H]=$-$1.16, $-$1.54, 
  and $-$2.02, respectively.

\begin{figure*}
    \includegraphics[width=13cm,trim={0.5cm 5.4cm 0.0cm 15.95cm},clip]{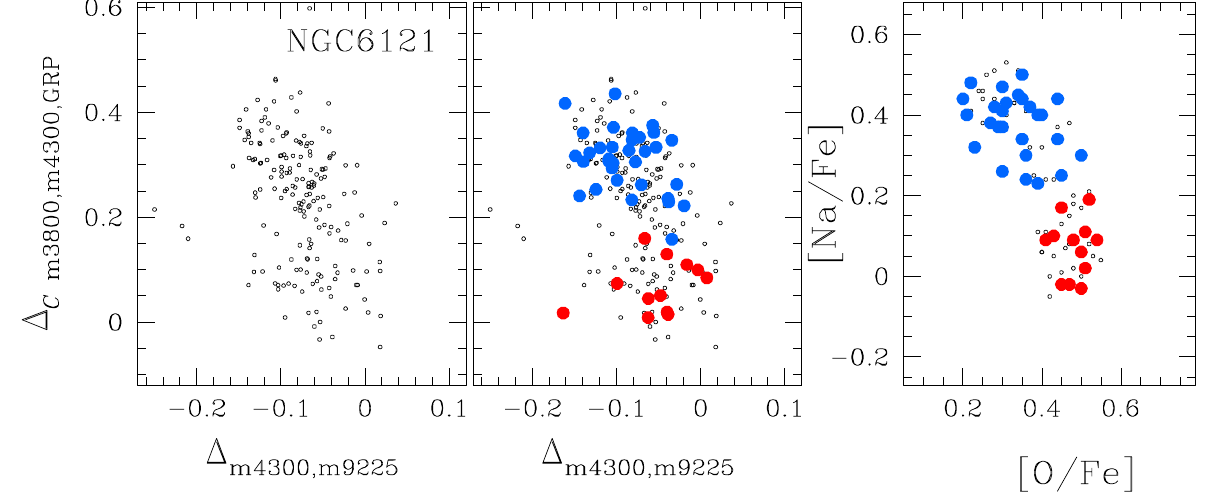}
    \includegraphics[width=13cm,trim={0.5cm 5.4cm 0.0cm 15.95cm},clip]{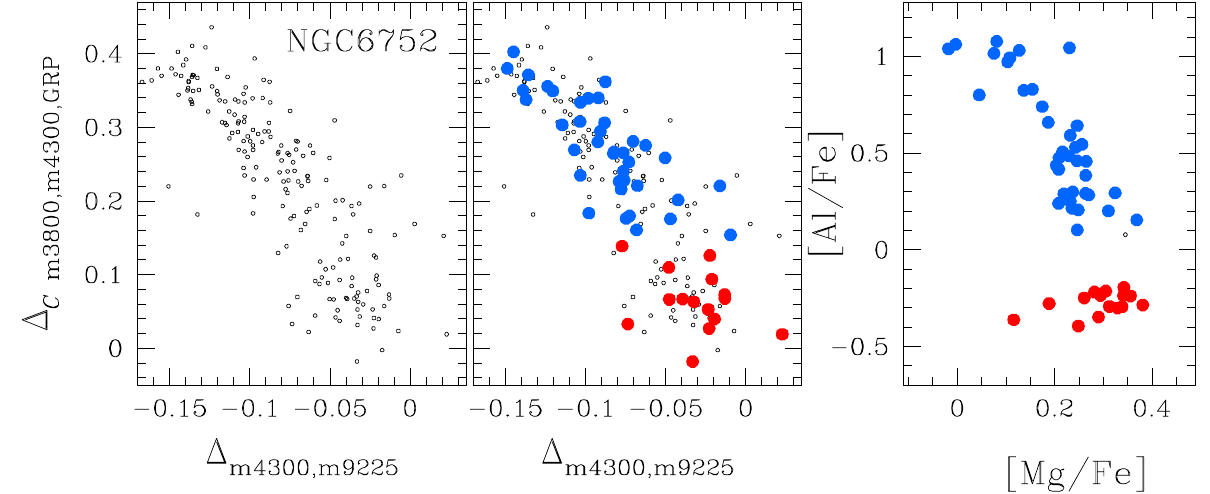}
        \includegraphics[width=13cm,trim={0.5cm 5.4cm 0.0cm 15.95cm},clip]{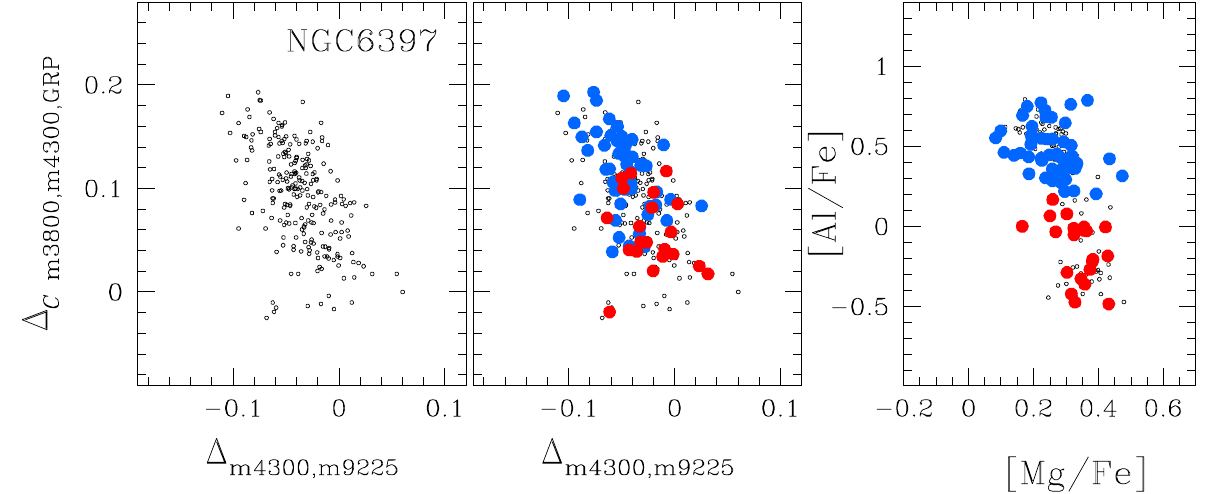}
   \caption{ $\Delta_{\rm m3800,m4300,GRP}$ vs.\,$\Delta_{\rm m4300,m9225}$ ChMs for RGB stars of NGC\,6121, NGC\,6752, and NGC\,6397 (left and middle panels). The right panels plot the anticorrelation between light elements. Specifically, we show [Al/Fe] vs.\,[Mg/Fe] for NGC\,6752 and NGC\,6397, and [Na/Fe] vs.\,[O/Fe] for NGC\,6121. The abundances of aluminum and magnesium are taken from the from the 17$^{\rm th}$ APOGEE data release, whereas sodium and oxygen abundances are derived by \citet{marino2008a}. 
    In the middle and right panels, the probable 1P and 2P stars, for which both stellar magnitudes and spectroscopic data are accessible and identified based on their chemical compositions, are shown in red and blue, respectively.
   }
    \label{fig:ChMsp}
\end{figure*}

 The resulting $\Delta_{\rm m3800,m4300,GRP}$ vs.\,$\Delta_{\rm m4300,m9225}$ ChMs of RGB stars are plotted in the left and middle panels of Figure\,\ref{fig:ChMsp}, whereas the right panels show the Na-O anticorrelation for NGC\,6121 \citep[from][]{marino2008a} and the Al-Mg anticorrelations for NGC\,6397, 
  and NGC\,6752 (from APOGEE DR17).
  The stars with both spectroscopic and photometric data are shown as large dots, with probable 1P and 2P stars, identified based on the Al-Mg or Na-O anticorrelations, colored red and blue, respectively.
 The gray points show the remaining stars.
 Similarly with what is observed for 47,Tucanae,  the majority of 1P stars in all clusters are concentrated near the origin of the ChM
  and have smaller average $\Delta_{\rm m3800,m4300,GRP}$  and larger $\Delta_{\rm m4300,m9225}$ absolute values than the 2P stars. 
 We found that the 1P and 2P stars of each cluster, identified from the ChM derived from the Gaia XP spectra populate distinct sequence of the  $\Delta_{\rm {\it C} U,B,I}$ vs.\,$\Delta_{\rm B,I}$ ChM. Moreover, we verified that there is correspondence between the stellar populations selected from the $\Delta_{C \rm F336W,F438W,F814W}$ vs.\,$\Delta_{\rm F438W,F814W}$  and the $\Delta_{C \rm F275W,F336W,F438W}$ vs.\,$\Delta_{\rm F275W,F814W}$ ChMs.
 
 These results indicate that the $\Delta_{\rm m3800,m4300,GRP}$ vs.\,$\Delta_{\rm m4300,m9225}$ ChM is an 
  effective tool for detecting 
  GC stars with different content of light-elements. 
  We derive similar conclusions by selecting the 1P and 2P stars from the [N/Fe] vs.\,[C/Fe] plane.
 
 The bulk of 1P and 2P stars populate distinct regions in the ChMs of NGC\,6121 and NGC\,6752. 
  On the contrary, the two stellar populations partially overlap each other in the ChM of NGC\,6397. This fact is due to the low metallicity of this GC and the moderate light-element  variations among its stars.

\begin{figure}
    \includegraphics[width=13cm,trim={0.5cm 5.4cm 0.0cm 7.0cm},clip]{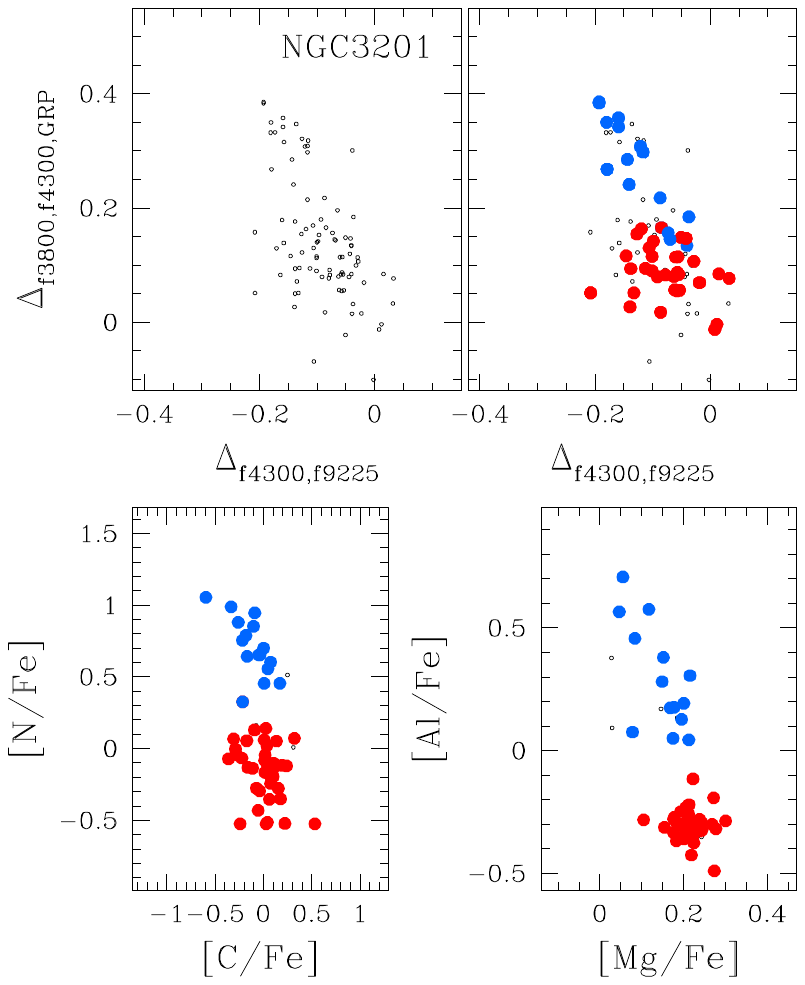}
   \caption{ $\Delta_{\rm m3800,m4300,GRP}$ vs.\,$\Delta_{\rm m4300,m9225}$ ChMs for RGB stars of NGC\,3201 (top panels). The  bottom panels show the Mg-Al and C-N anticorrelations from the 17$^{\rm th}$ APOGEE data release. The 2P stars with large abundances of aluminum and nitrogen, are colored blue, whereas the remaining stars with available photometry and spectroscopy are represented with blue dots.} 
    \label{fig:ChMspNGC3201AlN}
\end{figure}

\begin{figure*} 
    \includegraphics[width=14.5cm,trim={0.5cm 5.5cm 0.0cm 4.7cm},clip]{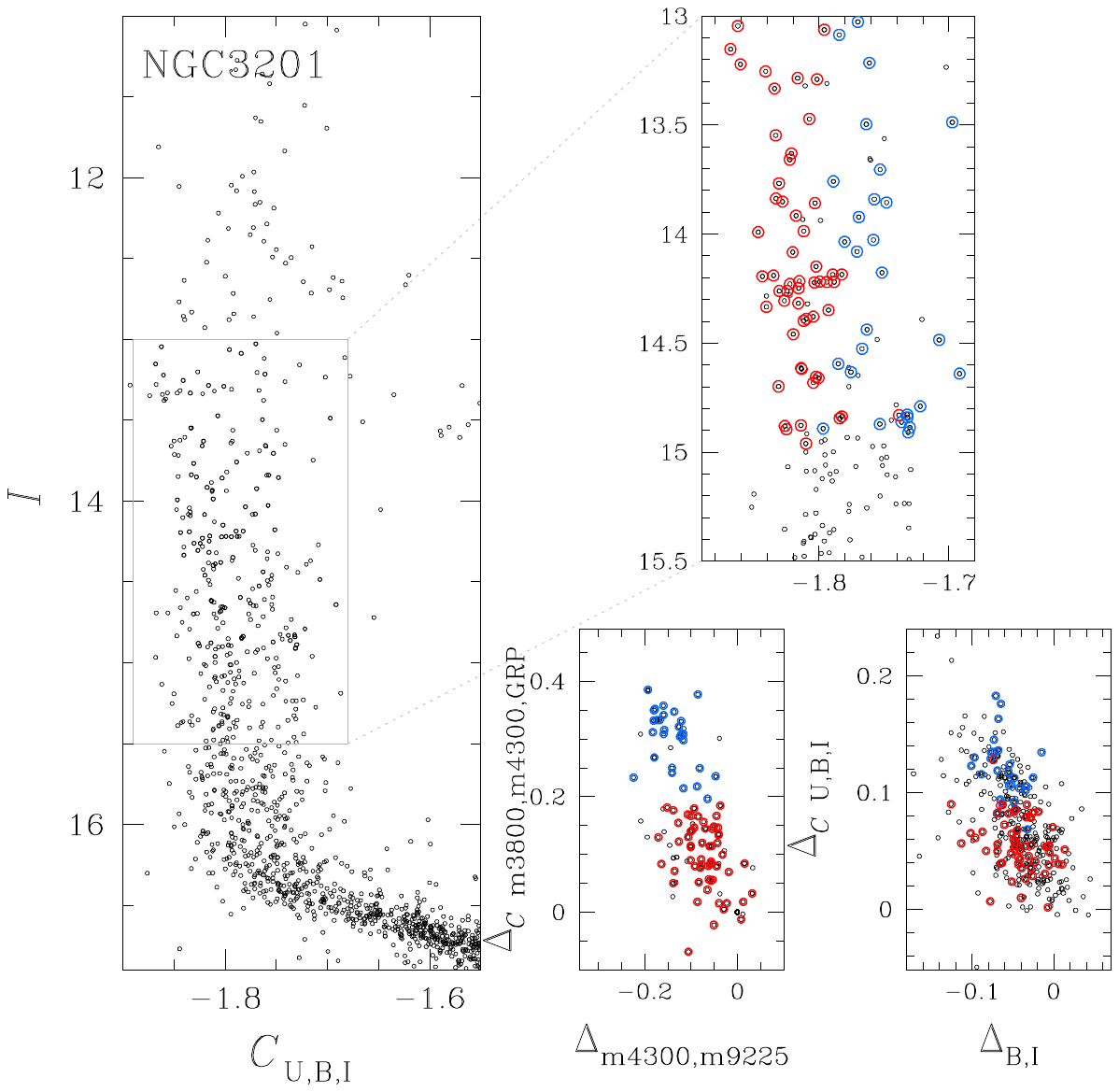}
    \caption{ $I$ vs.\,$C_{\rm U,B,I}$ pseudo-CMD of NGC\,3201 (left panel). The top-right panel is a zoom-in of the RGB region where the sequences of 1P and 2P stars are more evident for stars with radial distance from the cluster center larger than four arcmin. The bottom-right panels show the $\Delta_{\rm m3800,m4300,GRP}$ vs.\,$\Delta_{\rm m4300,m9225}$ (left) and the $\Delta_{C{\rm U,B,I}}$ vs.\,$\Delta_{\rm B,I}$ (right) ChMs. Probable 1P and 2P stars, selected from the ChM obtained from Gaia XP spectra, are colored red and blue, respectively.}
    \label{fig:Cubi3201}
\end{figure*}
\begin{figure*} 
    \includegraphics[width=12.5cm,trim={0.5cm 5.5cm 0.0cm 4.7cm},clip]{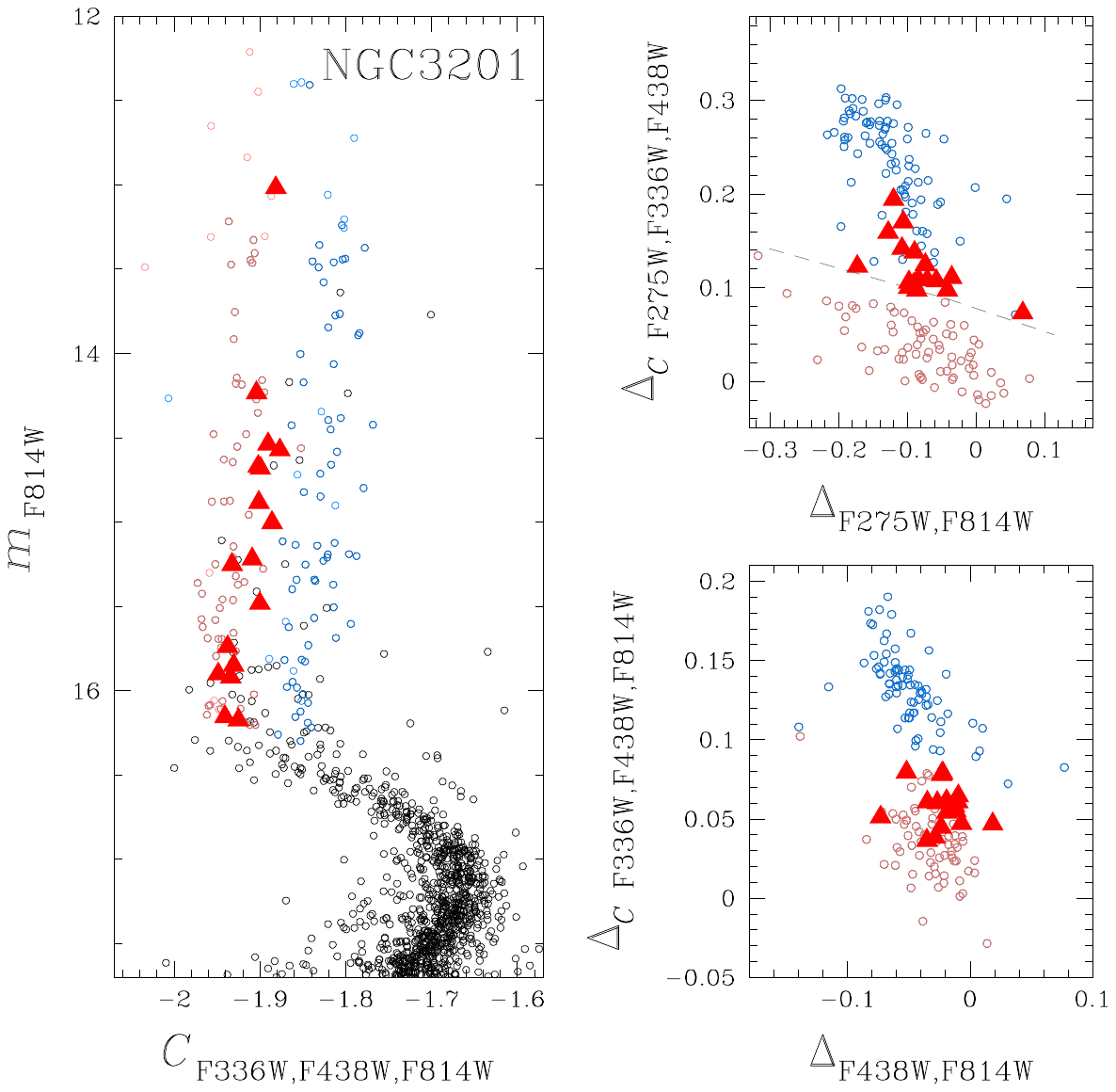}
    \caption{ $m_{\rm F814W}$ vs.\,$C_{\rm F336W,F438W,F814W}$ pseudo-CMD of NGC\,3201 (left). Right panels compare the $\Delta_{C \rm F275W,F336W,F438W}$ vs.\,$\Delta_{\rm F275W,F814W}$ (top) and the $\Delta_{C \rm F336W,F438W,F814W}$ vs.\,$\Delta_{\rm F438W,F814W}$ (bottom) ChMs of RGB stars. The dashed gray line in the top-right ChM separates the bulk of 1P and 2P stars identified by \citet{milone2017a}, which are colored red and blue, respectively, in all panels. The red triangles mark the 2P stars with low $\Delta_{C \rm F336W,F438W,F814W}$ values.}
    \label{fig:NGC3201hst}
\end{figure*}

\subsection{Multiple populations in NGC3201}\label{subsec:NGC3201}

 In the following, we investigate NGC\,3201 by using the same method adopted above for the other studied GCs.
The $\Delta_{\rm m3800,m4300,GRP}$ vs.\,$\Delta_{\rm m4300,m9225}$ diagram for RGB stars is shown in the top panels of Figure\,\ref{fig:ChMspNGC3201AlN} and reveals a group of stars with $\Delta_{\rm m3800,m4300,GRP} \gtrsim 0.2$ mag, alongside another group of stars with smaller $\Delta_{\rm m3800,m4300,GRP}$ values. 

The bulk of probable 2P stars, selected from the Al-Mg and the C-N anticorrelations, are colored red and correspond to the group of stars with large values of $\Delta_{\rm m3800,m4300,GRP}$ in the ChM, whereas the remaining stars in the ChM are Al-poor and N-poor.
 Noticeably, at odds with what we observed in 47\,Tucanae, NGC\,6121, and NGC\,6752, the group of ChM stars with low values of $\Delta_{\rm m3800,m4300,GRP}$ may exhibit intrinsic light-element variations, as suggested by the fact that the nitrogen abundance spans a range of $\sim$0.5 dex.
 
  To compare the ChMs from Gaia-based and ground-based photometry, we take advantage from the photometric catalogs by \cite{stetson2019a}.
  The photometry has been analyzed by following the same methods of our previous works \citep[e.g.\,][]{milone2018a, cordoni2020a, jang2022a}. In particular, we selected the cluster members from Gaia DR3 proper motions and parallaxes and corrected the photometry for the effects of differential reddening as in \cite{milone2012b}. The resulting $I$ vs.\,$C_{\rm U,B,I}$ is plotted in the left panel of Figure\,\ref{fig:Cubi3201} for the NGC\,3201 stars with radial distance larger than 1.5 arcmin. Moreover, in the bottom-right panels we compared the $\Delta_{\rm m3800,m4300,GRP}$ vs.\,$\Delta_{\rm m4300,m9225}$  and the $\Delta_{C{\rm U,B,I}}$ vs.\,$\Delta_{\rm B,I}$ ChMs \citep[][]{jang2022a}. For completeness, in the top-right panel we show the $I$ vs.\,$C_{\rm U,B,I}$ diagrams for RGB stars with radial distances larger than 4.0 arcmin from the cluster center.
   The probable 2P stars, selected from the $\Delta_{\rm m3800,m4300,GRP}$ vs.\,$\Delta_{\rm m4300,m9225}$ ChM, are colored blue, while the remaining star are plotted in red.
   As expected, the selected stellar groups occupy distinct sequences in both the ChMs and the pseudo-CMD, confirming that these diagrams consistently identify the same stellar populations.

Finally, Figure\,\ref{fig:NGC3201hst}, compares the $m_{\rm F814W}$ vs.\,$C_{\rm F336W,F438W,F814W}$ and $\Delta_{C \rm F336W,F438W,F814W}$ vs.\,$\Delta_{\rm F438W,F814W}$ diagrams, which are the {\it HST}-analogs of $I$ vs.\,$C_{\rm U,B,I}$ and $\Delta_{C \rm U,B,I}$ vs.\,$\Delta_{\rm B,I}$, with the $\Delta_{C \rm F275W,F336W,F438W}$ vs.\,$\Delta_{\rm F275W,F814W}$ diagram, the most commonly used ChM for studying multiple populations along the RGB.
The dashed line superimposed on the latter ChM is derived by \citet{milone2017a} and separates the bulk of 1P and 2P stars.
Clearly, all 1P stars (red circles) are located along the RGB sequence with low $C_{\rm F336W,F438W,F814W}$ values and cluster near the origin of the $\Delta_{C \rm F336W,F438W,F814W}$ vs.\,$\Delta_{\rm F438W,F814W}$ diagram. Moreover, all 2P stars (blue circles) have large values of $C_{\rm F336W,F438W,F814W}$ and $\Delta_{C \rm F336W,F438W,F814W}$.
However, we notice a significant number of 2P stars (red triangles) that populated the RGB sequence with low $C_{\rm F336W,F438W,F814W}$ values and cluster near the origin of the $\Delta_{C \rm F336W,F438W,F814W}$ vs.\,$\Delta_{\rm F438W,F814W}$ diagram.

We conclude that the available photometric diagrams based on $U,B,I$ photometry allow us to disentangle 2P stars of NGC\,3201 with large nitrogen variations (hereafter extreme 2P, or e2P stars), whereas the 2P stars that are mildly enhanced in nitrogen are mixed with the 1P. Due to the similar results obtained from $U,B,I$ Gaia-based photometry, the same conclusion can be extended to the $\Delta_{\rm m3800,m4300,GRP}$ vs.\,$\Delta_{\rm m4300,m9225}$ ChM.

\begin{figure*} 
    \includegraphics[width=12.5cm,trim={0.5cm 5.5cm 0.0cm 15.25cm},clip]{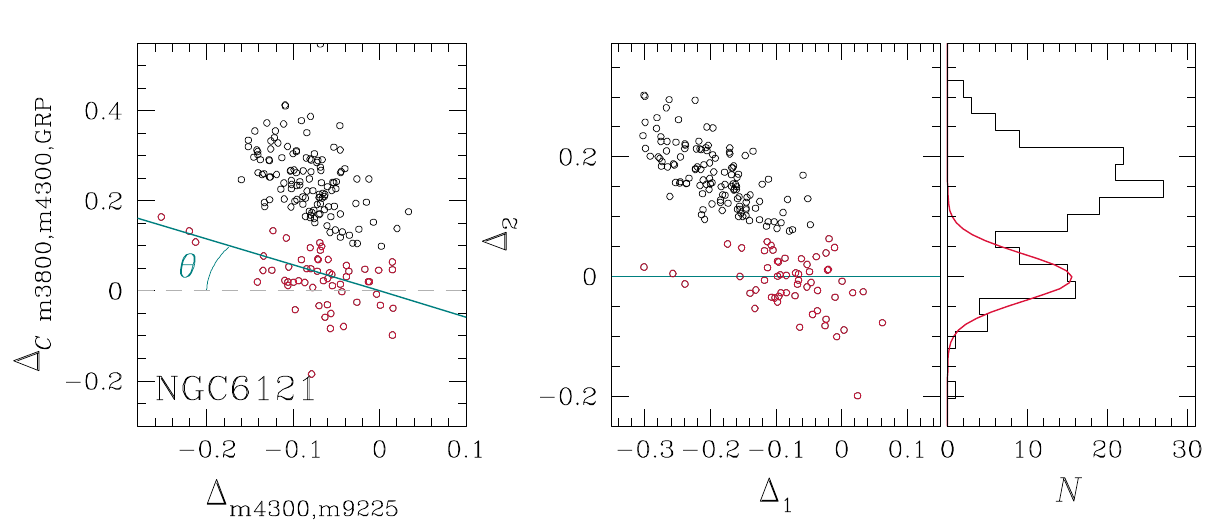}
    \caption{ Method used to measure the fractions of 1P and 2P stars in NGC\,6121. The left panel reproduces the ChM of Figure\,\ref{fig:ChMsp},  where we colored red the probable 1P stars. 
     The aqua line is the best-fit straight line through the red stars and forms an angle of $\theta=30^{o}$ with the gray dashed line.
    The middle panel displays the rotated $\Delta_{2}$ vs.\,$\Delta_{1}$ diagram, whereas the histogram distribution of  $\Delta_{2}$ is plotted in the right panel. The red curve is the best-fit Gaussian of the 1P stars}\label{fig:pratio}
\end{figure*}

\section{Fraction and radial distribution of multiple populations} \label{sec:RD}
 To estimate the fractions of 1P and 2P stars in each GC, we adopted the procedure illustrated in Figure\,\ref{fig:pratio} for NGC\,6121, which has been already used in previous work on GCs \citep{milone2017a}.
 The aqua line represents a linear fit to the 1P stars.  The $\Delta_{2}$ vs.\,$\Delta_{1}$ diagram plotted  on the middle of Figure\,\ref{fig:pratio} was generated by rotating counterclockwise the ChM by $\theta=30^{o}$, where $\theta$  is traced from 
 the green line to  the dashed horizontal line.
 We obtained the histogram distribution of $\Delta_{2}$ for the cluster members and displayed it in the right-hand panel.

The distribution of bona-fide 1P stars is modeled using a Gaussian curve through least squares fitting.
The fraction of 1P stars is estimated by taking the ratio of the area under the Gaussian curve (red line in Figure\,\ref{fig:pratio}) to the total area of the histogram.

 We found that the 1P of NGC\,6121 comprises 30.0$\pm$3.1\% of the total numbers of studied RGB stars. Similarly, we derived the fractions of 1P stars along the RGB of 47\,Tucanae, 
  and NGC\,6752, which correspond to 48.8$\pm$2.4, 
  and 25.9$\pm$3.2\%,  respectively. 
  In the case of NGC\,3201, where it is not possible to disentangle 1P stars and 2P stars with moderate light-element variations, we derived the fraction of e2P stars, which corresponds to 25.9$\pm$3.6\%.
  We excluded NGC\,6397 from the analysis because 
   1P and 2P stars are partially mixed in its ChM.

 To analyze the radial distribution of multiple populations within each cluster, we segmented our sample into groups of stars with equal populations at varying radial distances and calculated the fractions of 2P stars (or e2P stars for NGC\,3201) using the previously discussed method.
We adopted two radial bins for NGC\,3201, three bins for NGC\,6121 and NGC\,6752, and four bins for 47\,Tucanae.

\begin{figure}
    \includegraphics[width=8.5cm,trim={0.5cm 6.75cm 0.5cm 10.25cm},clip]{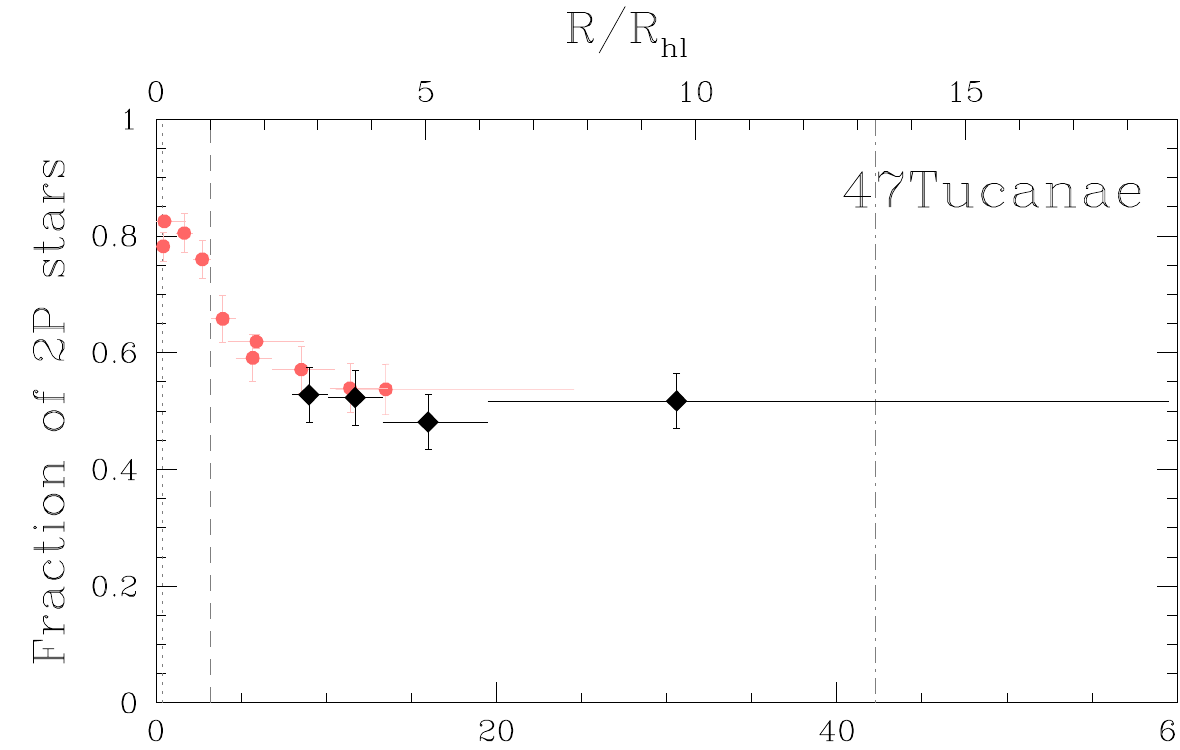}
    \includegraphics[width=8.5cm,trim={0.5cm 6.75cm 0.5cm 11.5cm},clip]{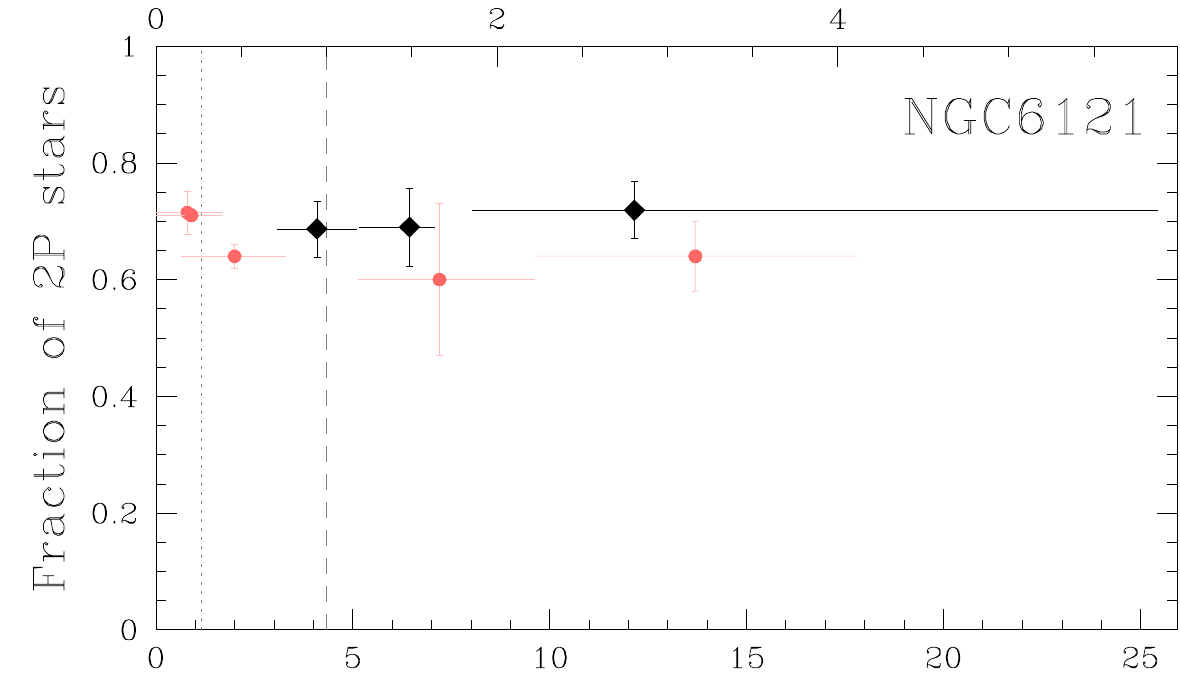}
    \includegraphics[width=8.5cm,trim={0.5cm 6.75cm 0.5cm 11.5cm},clip]{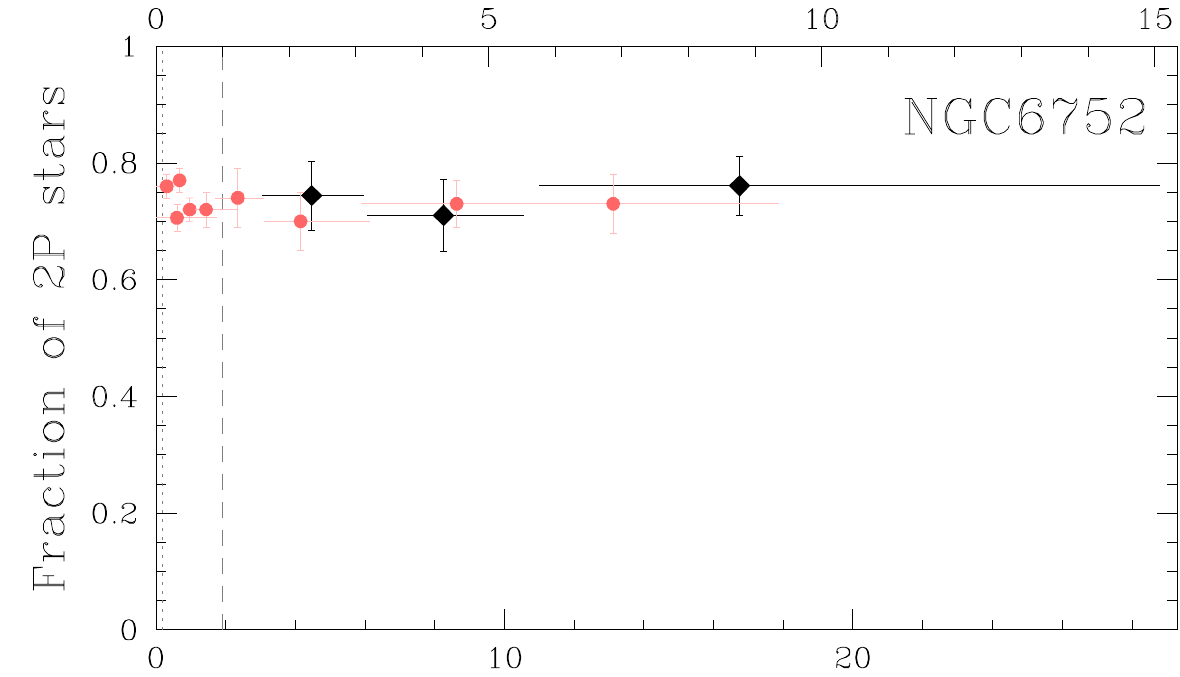}
    \includegraphics[width=8.5cm,trim={0.5cm 5.5cm 0.5cm 11.5cm},clip]{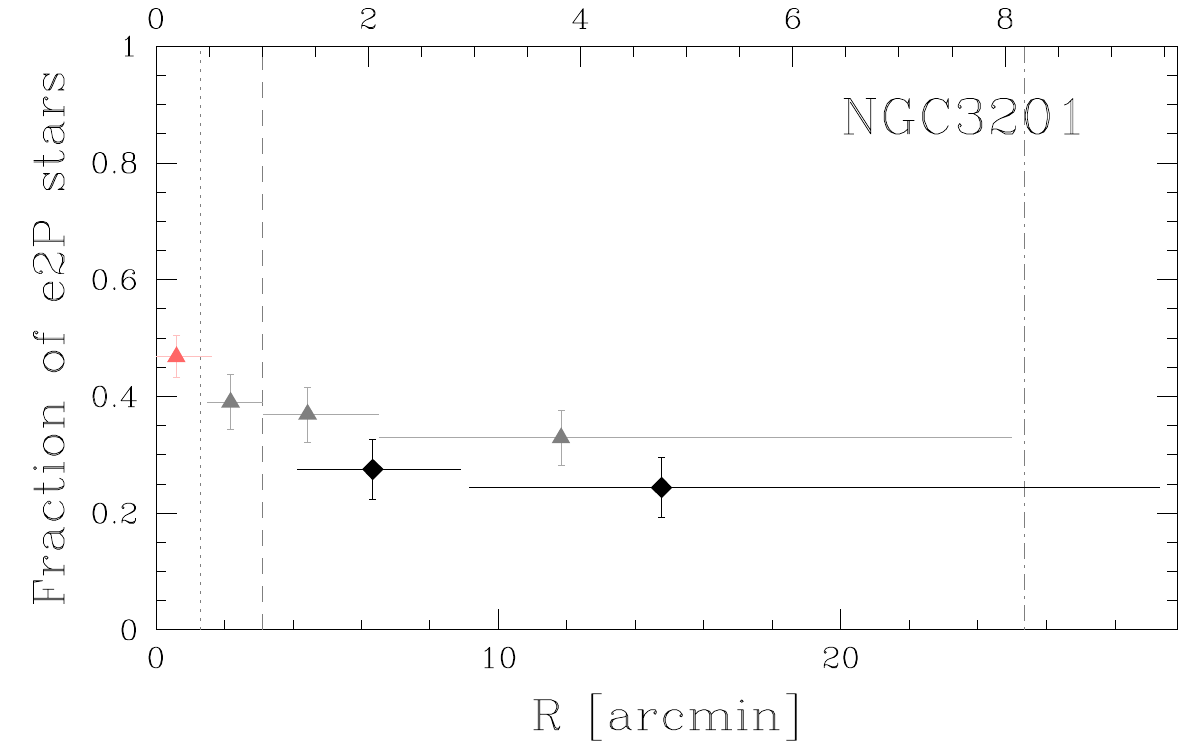}
    \caption{Fraction of 2P stars as a function of the radial distance from cluster center in 47\,Tucanae, NGC\,6121, and NGC\,6752, and fraction of e2P stars in NGC\,3201. The results   obtained from  the photometry derived from Gaia XP spectra are represented with black diamonds, whereas literature determinations \citep{milone2013a, milone2014a, milone2017a, milone2019a, nardiello2015a, dondoglio2021a, dondoglio2022a, marino2024a} are plotted with light-red dots.  The gray and light-red triangles show the fractions of e2P stars in NGC\,3201 derived in this paper from $U,B,I$ and $F336W,F438W,F814W$ photometry, respectively.
     The horizontal segments associated with each point mark the radial interval used to derive the corresponding fraction of 2P stars. The dotted, dashed, and dashed-dot lines indicate the core, half-light, and tidal radii, respectively, from the 2010 version of the \citet{harris1996a} catalog. }
    \label{fig:RD}
\end{figure}

The results are illustrated in Figure\,\ref{fig:RD}, 
 where we combined findings from this paper with those from previous studies, to analyze the radial behavior of the multiple populations. 
 
  In NGC\,3201, we also take advantage from the $\Delta_{C{\rm U,B,I}}$ vs.\,$\Delta_{\rm B,I}$ ChM to fully cover the field of view within the cluster tidal radius.
We used this diagram to compute the fractions of 2Pe stars in three different radial bins following the procedure illustrated in Figure\,\ref{fig:pratio} \citep[see also][]{jang2022a}.

 The fractions of 2P stars in both NGC\,6121 and NGC,6752 remain constant across the studied fields of view and are comparable to the values obtained from literature in the central regions, suggesting that their 1P and 2P stars are fully mixed.
In contrast, the fractions  of e2P and 2P stars that we measured in the external regions of NGC\,3201 and 47\,Tucanae,  respectively, are much smaller than those observed in the cluster centers, thus indicating that their second populations are more  concentrated toward the center compared to  the 1P.

\section{Summary and discussion}\label{sec:summary}

We used Gaia DR3 low-resolution XP spectra to introduce new photometric bands that are sensitive to the different chemical compositions of multiple populations in GCs. 
To do this, we used simulated spectra of RGB stars in 47\,Tucanae with the same metallicities but different light-element abundances as expected for 1P and 2P stars in GCs. 
We found that  
 diagrams based on photometry in the m3800$^{0200}$, m4300$^{0100}$, and m9225$^{0150}$ bands introduced in our work  and in the G$_{\rm RP}$ filter of Gaia DR3,   are able to efficiently 
  identify multiple populations among giant stars of 47\,Tucanae.In particular, the $\Delta_{\rm m3800, m4300,GRP}$ vs.\,$\Delta_{\rm m4300,m9225}$ ChM  
 enabled us to distinguish between 1P and 2P stars along the RGB, red HB, and AGB, as well as to identify metallicity variations among the 1P stars.

The photometric identification of 1P and 2P stars  is validated by the comparison of the ChM with the abundances of C, N, O, Mg, Al, and Na available from the literature and with the photometric diagrams constructed with the U,B,V,I magnitudes, which are traditionally used to identify  multiple populations in GCs.  
Similarly,  we clearly identified 1P  and 2P stars along the RGBs of NGC\,3201, NGC\,6121, and NGC\,6752. 
The sequences of 1P and 2P stars in the ChM of NGC,6397 partially overlap,  
 due to the small differences
 in the light-element abundances of its stars.

The ChMs introduced in this paper enable the determination of 1P and 2P star fractions in the outskirts of clusters, allowing the investigation of the radial behavior of multiple stellar populations near the tidal radius. The current distribution of 1P and 2P stars in GCs provides insights into their initial configurations and helps constrain models of GC formation.
There is widespread acceptance that in several GCs, such as 47\,Tucanae, NGC\,2808, NGC\,5272, and NGC\,5927, 2P stars exhibit a higher central concentration compared to 1P stars. In contrast, the radial distributions of multiple populations in some other clusters, including NGC\,6752, NGC\,6121, NGC\,6366 and NGC\,6838 remain consistent with one another \citep{ dondoglio2021a, dondoglio2022a, lee2019a, leitinger2023a, milone2012a,  milone2019a, nardiello2015a, simioni2016a}.
These observations align with scenarios in which  2P  stars originate in the cluster center and exhibit greater initial central concentration than the 1P stars. Consequently, clusters featuring persistently centrally concentrated 2P stars retain the imprint of the original stellar distribution.
In contrast, the 1P and 2P stars in other GCs become thoroughly mixed over time as a result of dynamical evolution \citep[e.g.][]{vesperini2013a, vesperini2018a, vesperini2021a}.
 
  This view has been challenged by recent studies, which suggest that in certain GCs, 2P stars are less centrally clustered than 1P stars. This conclusion implies that 1P stars might actually form with more central concentrations.
 NGC\,3201 stands out as a notable example illustrating this possibility \citep{leitinger2023a}.

  The ChMs that we derived from Gaia XP allow us to derive the fractions of 1P and 2P stars in the cluster outskirts, and explore a radial interval that is more extended than that analyzed in previous studies.
   Our findings, combined with literature results, confirm that the stellar populations of NGC\,6121 and NGC\,6752 share similar radial distributions. In contrast, 
    47\,Tucanae 
    presents a more centrally concentrated distribution of 2P stars compared to the 1P.
   The fraction of 2P stars declines from $\sim$80\% in the cluster core to $\sim$50\% at a radial distance of $\sim$3 half-light radii and remains nearly constant in the outermost cluster regions. These results are  
   consistent with the findings from previous studies \citep[e.g.][]{milone2012b, milone2019a, cordero2014a, lee2019a, lee2022a, dondoglio2021a, leitinger2023a}. 

  Similarly to what we observe in 47\,Tucanae, the e2P stars of NGC\,3201 are centrally concentrated. Their fraction ranges from $\sim$55\% in the cluster core to $\sim$25\% between the half-light and the tidal radius. 
  This result is in tension with the conclusions by \citet{leitinger2023a} that the 1P of  NGC\,3201 are more-centrally concentrated than 2P stars.
   Our findings are in qualitative agreement with those by \citet{cadelano2024a} for radial distances smaller than $\sim 4$ arcmin from the cluster center. However, we do not confirm the possibility of a bimodal radial distribution suggested by these authors with a fraction of e2P stars  
  larger than $\sim$70\% in its outermost regions.  
  
   Our results on the radial distribution of NGC\,6121, NGC\,6752, 47\,Tucanae, and NGC\,3201 are all consistent with the predictions of the scenarios for the GC formation  where the 2P stars originate in the innermost cluster regions.
   We do not confirm the possibility suggested by Leitinger and collaborators and based on their conclusions on NGC\,3201 that the 1P of some GCs can be more centrally  clustered than the 1P at the epoch of cluster birth. 
     On the contrary, the radial distribution of its multiple stellar populations is comparable with that of other dinamically young star clusters, where the 2P has a higher degree of central concentration  
     than the 1P.

\section*{acknowledgments} 
This work has been funded by the European Union – NextGenerationEU RRF M4C2 1.1 (PRIN 2022 2022MMEB9W: "Understanding the formation of globular clusters with their multiple stellar generations", CUP C53D23001200006), 
from INAF Research GTO-Grant Normal RSN2-1.05.12.05.10 -  (ref. Anna F. Marino) of the "Bando INAF per il Finanziamento della Ricerca Fondamentale 2022", and from the European Union’s Horizon 2020 research and innovation programme under the Marie Skłodowska-Curie Grant Agreement No. 101034319 and from the European Union – NextGenerationEU (beneficiary: T. Ziliotto).
\small
  
\section*{Data availability}
The data underlying this article will be shared upon reasonable request to the corresponding author.

\bibliography{ms}

\end{document}